\documentstyle[epsfig,12pt]{article}
\begin{document}
%
% **********************New command and definitions **************
%
%
\begin{titlepage}
\vspace{3 ex}
%
% ******************************* your title ************************
%
\begin{center}
{
\LARGE \bf \rule{0mm}{7mm}{\boldmath B Physics and Extra Dimensions}\\
}

\vspace{4ex}
% Your name and Institution. Please do not include street addresses
% For example; University of California Los Angeles, California 90024,
%U.S.A.
% or Centre d'Etudes Nucleaires-Saclay, 91191 Gif-sur-Yvette, France.
% would be sufficient.

{\large
J.F.Oliver, J.Papavassiliou, and A.Santamaria}
\vspace{1 ex}

{\em Departament de F\'{\i}sica Te\`orica 
and IFIC, Universitat de Val\`encia -- CSIC\\
E-46100 Burjassot (Val\`encia), Spain}

\vspace{2 ex}

\end{center}

\vspace{2 ex}
%
% ************************* Your abstract ***************************
%
\begin{abstract}

We compute
the dominant new physics contributions 
to the processes \( Z\rightarrow b\bar{b} \) and 
\( B \)-\( \bar{B} \) in the context
of two representative models with extra dimensions.
The main thrust of the calculations focuses on how to  
control the effects   
of the infinite tower of Kaluza-Klein modes inside 
the relevant one-loop diagrams. 
By comparing the results with  
the existing experimental data, most importantly   
those for \( R_{b} \), we show that 
one may derive interesting lower bounds on the size 
of the compactification scale \( M_{c} \).

\end{abstract}

\vspace{20 ex}

\centerline{\it {Talk given at the}}

\centerline{\it { 8th International Conference 
on B Physics at Hadron Machines -BEAUTY2002}}

\centerline{\it {Santiago de Compostela, Spain, 17-21 June, 2002}}

\centerline{\it{To appear in the Proceedings}}

\end{titlepage}

%*********************************************************
%
\setlength{\oddsidemargin}{0 cm}
\setlength{\evensidemargin}{0 cm}
\setlength{\topmargin}{0.5 cm}
\setlength{\textheight}{22 cm}
\setlength{\textwidth}{16 cm}
\setcounter{totalnumber}{20}
\clearpage\mbox{}\clearpage

\pagestyle{plain}
\setcounter{page}{1}
%
% **************** Start of text **********************************

\section{Introduction}
%\label{sec:intro}
In the last years there has been a revival of interest in models where
the  ordinary four  dimensional Standard  Model (SM)  arises as  a low
energy effective theory  of models living in five  or more dimensions,
with         the         extra        dimensions         compactified.
\cite{arkani-hamed:1998nn,arkani-hamed:1998rs,antoniadis:1990ew,
antoniadis:1994jp}.  Such models  arise naturally in string scenarios,
and  merit  a  serious  study,  mainly  because  of  the  plethora  of
theoretical and  phenomenological ideas associated with  them, and the
flexibility  they  offer  for  realizing new,  previously  impossible,
field-theoretic constructions.   Models with compact  extra dimensions
are  in general  not renormalizable,  and  one should  regard them  as
low-energy  manifestations  of  some  more  fundamental  theory.   The
effects  of  the  extra   dimensions  are  communicated  to  the  four
dimensional world through the presence of infinite towers of KK modes,
which modify qualitatively the  behavior of the low-energy theory.  In
particular,  the non-renormalizability  of  the theory  is found  when
summing  the infinite  tower of  KK states.   Tthe size  of  the extra
dimensions  can be  surprisingly large  without  contradicting present
experimental          data          (see         for          instance
\cite{antoniadis:1994yi,pomarol:1998sd,
antoniadis:1999bq,nath:1999mw,masip:1999mk,
delgado:1999sv,rizzo:1999br,carone:1999nz,nath:1999fs}).   This  opens
the  door to  the  possibility of  testing  these models  in the  near
future.   Most importantly,  the lowest  KK states,  if  light enough,
could be produced in the next generation of accelerators.

$B$ phenomenology can provide important 
generic tests for new physics 
(\cite{Stone:2001yn,Stone:2001jh} and references therein).   
For example,
the present experimental value of \( R_{b} \), 
\( R_{b}^{\mathrm{exp}}=0.2164\pm 0.00073 \)  
is perfectly compatible
with the standard model \cite{groom:1998in}, which predicts  
\( R_{b}^{\mathrm{SM}}=0.2157\pm 0.0002 \).  
Deviation due to new physics must be accomodated in this small 
window, a fact which furnishes non-trivial constraints on the possible 
models. 

In the SM the most important corrections are those
enhanced by the large top-quark mass in \( Z\rightarrow b\bar{b} \)
and  $B-\bar{B}$ mixing 
\cite{akhundov:1986fc,bernabeu:1988me,bernabeu:1991ws,beenakker:1988pv,
buchalla:1996vs,bernabeu:1997zh},
and \( \rho  \) parameter. 
In this talk we will show that, due to this same enhancement,
one may derive valuable information on the size of the
extra dimension(s), 
through the study of the one-loop contributions of the KK
modes to the processes $Z\to b \bar{b}$ 
and $B-\bar{B}$ mixing \cite{Papavassiliou:2000pq}.

\section{ $Z\to b \bar{b}$ and New Physics}

\begin{figure}[!t]
\begin{center}
\includegraphics[width=5cm]{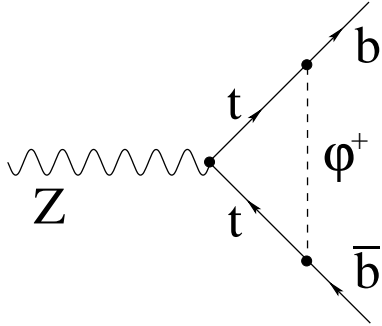}
\end{center}
\caption{\label{figure1} The leading contribution in the SM case}
\end{figure}

The effective \( Z\bar{b}b \) vertex is usually parametrized as 
\begin{equation}
\frac{g}{c_{W}}\, \bar{b}\gamma ^{\mu }(g_{L}P_{L}+g_{R}P_{R})bZ_{\mu }
\label{effvert}
\end{equation}
where \( P_{L}=(1-\gamma _{5})/2 \) and \( P_{R}=(1+\gamma _{5})/2 \) are,
respectively, the left and right chirality projectors. 
At one loop, the dominant corrections amount
to shifts in the coupling \( g_{L} \) and \( g_{R} \).
Thus, we will write \( g_{L}=-\frac{1}{2}+\frac{1}{3}s_{W}^{2}+\delta g^{\mathrm{SM}}_{L}+\delta g^{\mathrm{NP}}_{L} \)
which contains the SM tree level contribution, the SM one-loop contributions
\( \delta g^{\mathrm{SM}}_{L} \), and the contributions coming from new physics
\( \delta g^{\mathrm{NP}}_{L} \), 
and similarly for the right-handed couplings
\( g_{R}=\frac{1}{3}s_{W}^{2}+\delta g^{\mathrm{SM}}_{R}+\delta g^{\mathrm{NP}}_{R} \).
However, in general, \( g_{R} \) obtains sub-dominant corrections only (not
proportional to the top quark mass), in both the SM and new physics. 
The dominant SM contribution comes from the Goldstone boson diagrams running
in the loop (see Fig.~\ref{figure1}) and reads
\begin{equation}
\label{eq:deltasm}
\delta g_{L}^{\mathrm{SM}}\approx \sqrt{2}G_{F}m_{t}^{4}\, i\int \frac{d^{4}k}{(2\pi )^{4}}\frac{1}{(k^{2}-m_{t}^{2})^{2}k^{2}}=\frac{\sqrt{2}G_{F}m_{t}^{2}}{(4\pi )^{2}}
\end{equation}

A shift in the \( Zb\bar{b} \) couplings 
gives a shift in \( R_{b}=\Gamma _{b}/\Gamma _{h} \)
(here \( \Gamma _{b}=\Gamma (Z\rightarrow b\bar{b}) \) and \( \Gamma _{h}=\Gamma (Z\rightarrow \mathrm{hadrons}) \))
given by \begin{equation}
\label{eq:rb}
R_{b}=R_{b}^{\mathrm{SM}}\frac{1+\delta ^{\mathrm{NP}}_{bV}}{1+R_{b}^{\mathrm{SM}}\delta ^{\mathrm{NP}}_{bV}},
\end{equation}
 where \begin{equation}
\label{eq:deltavnp}
\delta ^{\mathrm{NP}}_{bV}=\frac{\delta \Gamma _{b}}{\Gamma _{b}^{\mathrm{SM}}}\approx 2\frac{g_{L}}{(g_{L})^{2}+(g_{R})^{2}}\delta g_{L}^{\mathrm{NP}}\approx -4.6\, \delta g_{L}^{\mathrm{NP}}.
\end{equation}
 gives the shift in \( \Gamma _{b} \) due to vertex corrections coming from
new physics, \( \Gamma _{b}=\Gamma ^{\mathrm{SM}}_{b}+\delta \Gamma _{b} \).
Here, quantities with the superscript ``SM'' 
represent standard model quantities including
complete radiative corrections. Note that non-vertex corrections are universal
for all quarks and cancel in the ratio \( R_{b} \).

\section{A model with fermions on the brane}

Extra dimensions may  or may
not be accessible  to all known fields, depending  on the specifics of
the underlying, more fundamental theory. Here we consider  
the simplest generalization of the SM, 
the so-called 5DSM, with fermions living in 
four dimensions, and gauge bosons
and a single scalar doublet propagating in 
five dimensions \cite{pomarol:1998sd}.
The relevant pieces of the five dimensional Lagrangian are (\( \mu =0,1,2,3 \)
are four dimensional indices and \( M=0,1,2,3,5 \)
are five dimensional ones) 
\begin{equation}
L=\int d^{5}x\left( \partial _{M}\varphi ^{\dagger }\partial ^{M}\varphi -\left( \bar{Q}_{L}Y_{u}u_{R}\, \varphi \, \delta (x^{5})+\mathrm{h}.\mathrm{c}.\right) +\cdots \right) 
\label{L5}
\end{equation}
 where \( \varphi (x^{M}) \) is the \( SU(2) \) Higgs doublet which lives
in five dimensions. \( Q_{L}(x^{\mu }) \) and \( u_{R}(x^{\mu }) \) are the
standard left-handed quark doublet and right-handed singlet, respectively, 
which
live in four dimensions (brane); this is enforced by the presence of the 
\( \delta (x^{5}) \).
We assume that \( x^{5} \)
is compactified on an orbifold $S^{1}/{Z}_{2}$.
Fields even under the ${Z}_{2}$ symmetry have zero modes
and are present in the low energy theory, whereas 
fields that are odd 
have only KK modes and disappear from the low energy spectrum. One
chooses the Higgs doublet to be even under the aforementioned symmetry
in order to have the standard Higgs boson. 
Fourier-expanding the Higgs field 
as 
\begin{equation}
\varphi (x^{\mu },x^{5})=\sum _{n=0}^{\infty }  \varphi _{n}(x^{\mu })
\cos \frac{nx^{5}}{R} \, ,
\end{equation}
substituting 
in the fifth dimensional Lagrangian, and integrating on the
fifth component, we obtain the following four dimensional 
Lagrangian for the KK modes \( \varphi _{n}(x) \):
\begin{eqnarray}
\mathcal{L} & = & \partial _{\mu }\varphi _{0}^{\dagger }\partial ^{\mu }\varphi _{0}-\left( \bar{Q}_{L}Y_{u}u_{R}\varphi _{0}+\textrm{h}.\textrm{c}.\right) \nonumber \\
 & + & \sum _{n=1}^{\infty }\left( \partial _{\mu }\varphi _{n}^{\dagger }\partial ^{\mu }\varphi _{n}-\frac{n^{2}}{R^{2}}\varphi _{n}^{\dagger }\varphi _{n}-\left( \bar{Q}_{L}Y_{u}u_{R}\sqrt{2}\varphi _{n}+\textrm{h}.\textrm{c}.\right) \right) \label{eq:lagrangian4d} 
\end{eqnarray}
The additional factor \( \sqrt{2} \) comes from
the normalization of the zero mode in the Fourier series. 

\begin{figure}[!t]
\begin{center}
\includegraphics[width=5cm]{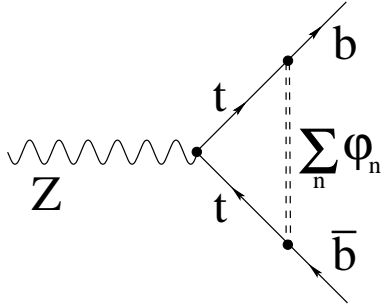}
\end{center}
\caption{\label{figure2} The KK tower corresponding to the Higgs doublet 
living in 5-D}
\end{figure}

As in the SM case, 
the dominant contributions to \( Z\rightarrow b\bar{b} \) for
large \( m_{t} \) come from the diagram Fig.~\ref{figure2} where
charged KK modes are running in the loop. 
The zero mode provides just the standard
model contribution due to the exchange of the 
Goldstone boson while the exchange
of the KK modes of the charged components of the Higgs doublet 
will give an
extra contribution. 
Since the coupling is universal summing all these contributions
amounts to replacing the propagator of the Goldstone boson
(in the Euclidean)
\begin{equation}
\frac{1}{k^{2}_E}\rightarrow \frac{1}{k^{2}_E}+
2\sum _{n=1}^{\infty }\frac{1}{k^{2}_E + n^{2}/R^{2}}=
\sum _{n=-\infty }^{\infty }\frac{1}{k^{2}_E + n^{2}/R^{2}}=
\pi R\frac{\coth (k_E\pi R)}{k_E}.
\label{effprop}
\end{equation}
This effective propagator  reduces for small \( k_E \) 
to the standard Goldstone propagator.
However,
for large \( k_E \) it goes as \( 1/k_E \), 
which means that the ultraviolet behavior
of this theory is worse than in the SM by one power of \( k_E \). 
Even though this propagator provides a convergent result 
for this particular diagram, 
its soft high energy behaviour
will trigger  eventually the non-renormalizability of the theory. 

Adding the KK modes we obtain \begin{equation}
\label{eq:deltanp}
\delta g^{\mathrm{NP}}_{L}\approx \delta g_{L}^{\mathrm{SM}}\left( F(a)-1\right) ,
\end{equation}
 where \( a=\pi Rm_{t} \), and \begin{equation}
\label{eq:integral}
F(a)=2a\int _{0}^{\infty }dx\frac{x^{2}}{(1+x^{2})^{2}}\coth (ax)
\end{equation}
 is the ratio of the non-standard/standard integrals. 
Expanding for small
\( a \) we obtain \begin{equation}
\label{eq:f-function}
F(a)\approx 1+a^{2}\left( -\frac{1}{3}-\frac{4}{\pi ^{2}}\zeta '(2)-\frac{2}{3}\log (a/\pi )\right) \approx 1+a^{2}\left( 0.80979-\frac{2}{3}\log (a)\right) 
\end{equation}
 where \( \zeta ' \) is just the derivative of the Riemman Zeta function.
Contributions from the KK towers of gauge bosons
running in the loop are not enhanced by the
top quark mass and are suppressed 
by factors \( (m_{W}/m_{t})^{2} \) with respect
to the contribution considered. Although such 
contributions are important for obtaining
the standard model value \cite{bernabeu:1988me} at the required precision, 
they can be neglected when estimating bounds on new physics. 

One finds \( F(a)-1=-0.24\pm 0.31 \); since $F(a)$
is always larger than 1, 
translating this result into an upper bound on \( F(a)-1 \) is 
subtle.
For this purpose we used the prescription of ref.~\cite{feldman:1998qc} and
found the 95\% CL limit of \( F(a)-1<0.39 \), which, after evaluation
of the integral of Eq.(\ref{eq:integral}), 
translates into an upper bound on \( a \), yielding \( a<0.56 \).
This amounts to a 
lower bound $M_{c}>0.98\: \mathrm{TeV}$  
on the compactifications scale \( M_{c}=1/R \), which is  
comparable to those obtained
from tree level processes \cite{nath:1999mw,masip:1999mk,delgado:1999sv,rizzo:1999br,carone:1999nz,nath:1999fs}. 

\section{Models with universal extra dimensions}

Scenarios  where all  SM
fields  live   in  higher  dimensions  have also been considered
(see for example  \cite{carone:1999nz,appelquist:2000nn});
we  will  refer  to this type  of  extra
dimension(s) as ``universal''.
From the  phenomenological point  of view,
the most  characteristic feature of such theories  is the conservation
of the KK  number at each elementary interaction  vertex 
\cite{carone:1999nz,appelquist:2000nn}.
As a result,
and contrary to  what happens in the non-universal  case, the coupling
of any excited (massive) KK mode to two zero modes is prohibited.  
This fact alters profoundly their production possibilities:
using normal  (zero-mode)  particles as  initial
sates, such modes cannot be resonantly produced, nor 
can a single KK mode appear in the final states, but  
must be  {\it pair-produced}.
In addition, the conservation of the KK number  
leads to the appearance of heavy stable
(charged   and  neutral)  particles,   which  may   pose  cosmological
complications (e.g. nucleosynthesis) \cite{appelquist:2000nn}; 
however, one-loop effects may overcome such problems 
\cite{Servant:2002aq}. 

\begin{figure}[!t]
\begin{center}
\includegraphics[width=5cm]{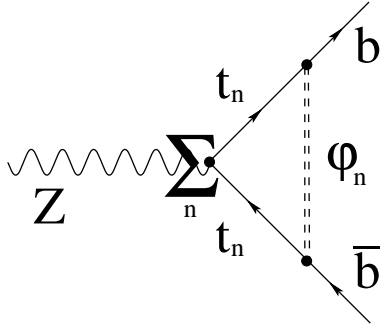}
\end{center}
\caption{\label{figure3} The sum goes over all particle in the loop 
which live in 5-D}
\end{figure}

It turns out that  in  the universal  case
the process $Z\to  b\bar{b}$ 
furnishes a less stringent bound compared to the  non-universal one
\cite{appelquist:2000nn}. 
The main difference now is that 
in the relevant Feynman graph one should sum over  
{\it all} particles inside the loop, Fig.~\ref{figure3}. The end result 
of the calculation is a lower bound for $M_c$ of 
about 300  GeV.

\section{\protect\( B\protect \)-\protect\( \bar{B}\protect \) mixing }

In the SM, the mixing between the \( B^{0} \) meson and its
anti-particle is also completely dominated by the top-quark contribution. The
explicit \( m_{t} \) dependence of the box diagrams is given by the 
loop function \cite{buchalla:1996vs}
\begin{equation}
S(x_{t})_{\mathrm{SM}}=
\frac{x_{t}}{4}\left[ 1+\frac{9}{1-x_{t}}-\frac{6}{(1-x_{t})^{2}}-\frac{6x_{t}^{2}\log (x_{t})}{(1-x_{t})^{3}}\right] \; ,\qquad x_{t}\equiv \frac{m^{2}_{t}}{M_{W}^{2}},
\end{equation}
 which contains the hard \( m_{t}^{2} \) term, i.e. \( x_{t}/4 \), induced
by the longitudinal \( W \) exchanges. The same function controls the top--quark
contribution to the \( K \)-\( \bar{K} \) mixing parameter \( \varepsilon _{K} \).
The measured top-quark mass, \( m_{t}=175 \)~GeV, 
implies \( S(x_{t})_{\mathrm{SM}}\sim 2.5 \).

\begin{figure}[!t]
\begin{center}
\includegraphics[width=5cm]{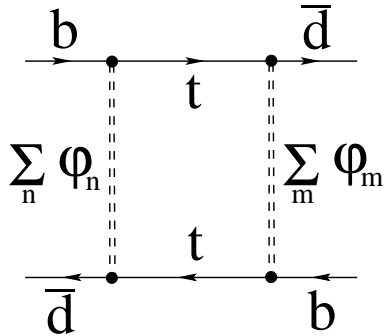}
\end{center}
\caption{\label{figure4} The box in the presence of KK corrections}
\end{figure}

Returning to the 5DSM model considered in section 3,
the KK modes of the charged components of the doublet also contribute to the
box diagram of Fig.~\ref{figure4}.
The total dominant contribution, SM plus KK modes, can be obtained
by substituting the SM scalar propagator by the effective one of 
Eq.~(\ref{effprop}). 
However, since this
modified propagator behaves as \( 1/k_{E} \) for large \( k_{E} \), and therefore,
the insertion of two propagators of this type turns the
box diagram into UV divergent. 
We write the correction to \( S(x_{t}) \) as 
\begin{equation}
S(x_{t})=S(x_{t})_{\mathrm{SM}}+\delta S(x_{t})\, ,\qquad \quad \delta S(x_{t})=\frac{x_{t}}{4}\, \left( G(a)-1\right) \, \, \, \, ,
\end{equation}
 where \( G(a) \) is again the 
ratio of the non-standard to standard box integrals, i.e. 
\begin{equation}
G(a) =  
2a^{2}\int _{0}^{\infty }dx\frac{x^{3}}{(1+x^{2})^{2}}\coth ^{2}(ax) \,; 
\label{eq:integral-box} 
\end{equation}
it is clearly divergent for \( x\rightarrow \infty  \). 

To evaluate $G(a)$ we cut off the
integral at \( x\approx n_{s}/a \), where \( n_{s} \) is related to the scale
at which new physics enters to regulate the five dimensional theory. In particular,
\( M_{s}\sim n_{s}M_{c} \) and \( n_{s}\gg 1 \). 
We obtain
\begin{equation}
G(a)\approx 1+a^{2}\left( -1.14314-\frac{4}{3}\log (a)+2\log (n_{s})\right) .
\end{equation}
For moderate values of \( a\sim 0.2 \) and \( n_{s}\sim 10 \) the new physics
correction is only about \( 0.2 \). For more extreme values (for instance \( a\sim 0.6 \)
and \( n_{s}\sim 100 \)), the corresponding contribution 
to \( G(a) \) is about \( 3 \). 
Notice also that the presence of diagrams with
gauge boson KK modes could modify the bounds on \( M_{c} \) by a factor of
about 20\%. However, given the uncertainty in the calculation of the 
box diagrams
due to the dependence on the scale \( M_{s}, \) estimating such effects seems
superfluous. The important point, however, is that the contribution from extra
dimensions to the function \( S(x_{t}) \) is always positive.

We can use the measured \( B^{0}_{d} \)-\( \bar{B}^{0}_{d} \) mixing to infer
the experimental value of \( S(x_{t}) \) and set 
a limit on \( \delta S(x_{t}) \).
The explicit dependence on the quark--mixing
parameters can be resolved by combining the constraints from \( \Delta M_{B^{0}_{d}} \),
\( \varepsilon _{K}, \) and \( \Gamma (b\rightarrow u)/\Gamma (b\rightarrow c) \).
In ref.~\cite{bernabeu:1997zh} a complete analysis of the allowed values for
\( S(x_{t}) \) was performed by varying all parameters in their allowed regions.
The final outcome of such an analysis is that \( S(x_{t}) \) could take values
within a rather large interval, namely 
$ 1<S(x_{t})<10 $.
Since most of the errors come from uncertainties in theoretical calculations,
it is rather difficult to assign confidence levels to the bounds quoted above.
The lower limit is very stable under changes of parameters, while the upper
limit could be modified by a factor of 2 by 
simply doubling some of the errors. 

Given that the standard model value for \( S(x_{t}) \) 
is \( S(x_{t})_{\mathrm{SM}}=2.5 \),
\emph{positive} contributions can be comfortably accommodated, whereas negative
contributions are more constrained. As we have seen, extra dimensions result
in \emph{positive} contributions to \( S(x_{t}) \); in fact one can obtain
values that could approach the upper limit of \( S(x_{t}) \) only for rather
small values of the compactification scale \( M_{c} \) and large values of
the scale of new physics, \( M_{s} \). It seems therefore that, at present,
the above bounds do not provide good limits on \( M_{c} \). On the other hand,
if future experiments combined with theoretical improvements were to furnish
a value for \( S(x_{t}) \) exceeding that of the SM, our analysis shows that
such a discrepancy could easily be accommodated in models with large extra dimensions. 

In conclusions, we have seen that the existing results from 
the processes $Z\to b \bar{b}$ and
$B-\bar{B}$ mixing can furnish valuable 
lower bounds on the size of the 
possible extra dimenions. These bonds are 
model-dependent, varying between 0.3 GeV to 1 TeV, a fact which 
corroborates the expectation that 
the corresponding KK states should be well within the 
reach of the LHC. 

\medskip

{\bf {Acknowledgments}}:
This research was supported by CICYT, Spain, under Grant AEN-99/0692,
in part by MCYT under the Grant BFM2002-00568, and
in part by the OCYT of  the Generalitat Valenciana 
under the Grant GV01-94.
JP thanks the organizers of BEAUTY 2002 for providing a very pleasant
atmosphere.

%***************** Bibliography ************************************
%\newpage

\end{document}